\documentclass[fleqn,a4paper,12pt]{article}
\usepackage{amssymb}
\usepackage{makeidx}
\usepackage{amsmath}
\usepackage{graphicx}
\usepackage{lscape}
\usepackage{a4}
\usepackage{epsfig}
\begin{document}

\thispagestyle{empty}
\newcommand{\al}{\alpha}
\newcommand{\bt}{\beta}
\newcommand{\s}{\sigma}
\newcommand{\lbd}{\lambda}
\newcommand{\vp}{\varphi}
\newcommand{\va}{\varepsilon}
\newcommand{\gm}{\gamma}
\newcommand{\G}{\Gamma}
\newcommand{\p}{\partial}
\newcommand{\om}{\omega}
\newcommand{\be} {\begin{equation}}
\newcommand{\lo}{\left(}
\newcommand{\ro} {\right)}
\newcommand{\ee} {\end{equation}}
\newcommand{\ba} {\begin{array}}
\newcommand{\ea} {\end{array}}
\newcommand{\ds}{\displaystyle}

\begin{center}
{\large \bf RELATIVISTIC WAVE EQUATIONS FOR INTERACTING, MASSIVE PARTICLES WITH
ARBITRARY HALF-INTEGER SPINS}\\
\vspace{4mm}
{J. Niederle}\\
{Institute of Physics of the Academy of Sciences of the Czech Republic,\\
Na Slovance 2, 18221 Prague 8, Czech Republic}\\

\vspace{3mm}

{A.G. Nikitin}\\
{Institute of Mathematics of the National Academy of Sciences of the Ukraine,%
\\
Tereshchenkivs'ka Street 3, 01601 Kiev, Ukraine}

\end{center}

\begin{abstract}
New relativistic wave equations (RWE) for massive  particles with
arbitrary half-integer spins $s$ interacting with external electromagnetic fields are proposed. 
They are based on wave functions which are irreducible tensors of rank $2n$ ($n=s-\frac12$) 
antisymmetric w.r.t. $n$ pairs of indices, whose components are bispinors. The form of RWE 
 is straightforward and free of
inconsistencies associated with the other approaches to equations describing interacting 
higher spin particles.
\end{abstract}







\section{ Introduction}

Over the years many relativistic wave equations (RWE) for description of
particles with arbitrary spin $s$ have been proposed and studied in detail
by the field- or group-theoretical methods (see e.g. \cite{1} -\cite{11} and
surveys in \cite{12} -\cite{14}). It turns out that the various proposed RWE
are more or less equivalent as far as free particles are concerned but
differ essentially in the physically more relevant cases, i.e., whenever
interactions of particles with an external electromagnetic or other field
are taken into account. In fact it has been discovered that several
difficulties arises for RWE describing higher spin particles interacting
with external fields. They are related to several mutually dependent facts
and can be briefly summarized as follows.

\begin{itemize}
\item  First, the wave function which is a solution of a given first order
RWE describing a particle with higher spin $s$ ($s>1/2$) must necessarily
have more components than are theoretically required (i.e., more than $%
2(2s+1)$). Hence the RWE should be provided with the appropriate number of
constraints to ensure the right number of independent components of the wave
function. While this requirement can be met in the case of the RWE for free particles, the
introduction of interactions with an external electromagnetic field may cause a failure in this 
respect. It leads either to too many constraints on the components of wave function or 
to not enough of them (for details see \cite{15}, \cite{16} or \cite{17}), or it yields 
to an unacceptable restriction on the external field already
discussed by Fierz and Pauli in \cite{3} (see also \cite{16}). The algebraic
criteria which determine whether or not the above mentioned difficulties will
arise, can be found in \cite{10}.

\item  Second, the wave function describing a higher spin particle interacting with external fields
 can propagate acausally since the corresponding
RWE may not be hyperbolic or the propagation speed of the wave function can be
lager then that of the velocity of light in the vacuum. This phenomenon
which was first discovered to the surprise of theoreticians by Velo and
Zwanziger \cite{15} in 1969 (see however also paper by Johnson and Sudarshan of \cite{22}) reopened the
problem of RWE once again - the problem that, after the papers of Salam and
Mathews \cite{18} and by Schwinger \cite{19} had been considered as
completely solved.

\item  Third, unacceptable changes in the anticommutation
rules for field components can occur when interactions with an
electromagnetic (or other)  field  are introduced \cite{22} (see also \cite{13}).

\item  In the fourth place, modes of complex frequency (i.e., the complex energy levels)
may appear for a system of higher spin particle interacting with a strong
external magnetic field (for details see \cite{20}).

\item  Fifth, starting with spin $s$ RWE for a free particle and
introducing to it interactions via minimal coupling
a charged particle is described whose gyromagnetic ratio $g$ is equal to ${\frac %
1s}$ instead of the desired $g=2$ (see, e.g., \cite{8},). The other  inconsistency of RWE with 
minimal interaction consists in the
absence of spin-orbit coupling \cite{23A}

\item  Let us remark that the above mentioned difficulties are in addition
to those which appear already in the free particle theory, namely,
that the charge of integer spin particle and  energy of half-integer spin 
particle are indefinite (see, e.g., \cite{7}, \cite{27}).
\end{itemize}

In order to complete this brief survey let us mention main disadvantages 
of the most frequently used approaches.

In the Bargman-Wigner formulation \cite{5} in which the wave function has $2s$
bispinorial indices and satisfies the Dirac equation for each of them the
main disadvantage consists in the impossibility to introduce
a minimal interaction because the resultant equations have trivial solutions only.
 The same is true for covariant
systems of equations proposed by Bakri \cite{28}.

In the Bhabha approach \cite{29} the corresponding equations admit the minimal interaction 
\cite{30}. But these equations describe multiplets of
particles with spins equal to $s,s-1,...,s_{0}$ where $s_{0}={\frac{1}{2}}$ or $%
s_{0}=0$ for half-integer and integer spins respectively.

The Lomont-Moses \cite{31}, Hagen-Hurley \cite{32}, and Dirac-like equations
with differential constraints \cite{33} are causal in the case of
anomalous interaction, but yield complex energies
for a particle interacting with  crossed electric and magnetic fields 
\cite{14}.

The Weinberg equations \cite{34} for particles of spin $s$ contain time
derivatives of order $2s$, and, as a result, admit nonphysical solutions.
For the recent analysis of these equations see ref. \cite{35}.

The relativistic Schr\"{o}dinger equations without redundant components \cite
{36} admit reasonable quasirelativistic approximations \cite{14},  however,
make troubles to  introduce minimal interaction since they are formulated in terms 
of integro-differential operators.

These inconsistencies of RWE for particles with higher spin $s$ are
especially provoking due to the following well-known experimental
facts: (i) that many baryonic resonances with spins equal to $s={\frac{1}{2}}%
,{\frac{3}{2}},...$ up to ${\frac{13}{2}}$ have been found and are well
established \cite{24}, (ii) that relatively stable and massive vector bosons 
$W^{+}$ and $W^{-}$ mediating weak interactions were discovered and has been
studied in great details, (iii) that there exists a number of composite
systems (e.g., exotic atoms \cite{omega}, or excited states of the Helium nuclei  ) whose
energy states and other properties should be described by the RWE for
particles of higher spin.

Moreover, in connection with the idea of unification of fundamental particle
interactions and of quantum theory with gravity in contemporary particle
physics (i.e., in string theories, supergravity, M-theory, etc.) many
interacting higher spin particles or other objects (p-branes) have been
introduced and must be consistently described (and not only in 3+1 dimensions!).

In the present paper we propose new equations for charged {\it massive} particles
with arbitrary half-integer spins interacting with an electromagnetic (or
other) external field. In fact we propose two kinds of models: one for a
single interacting particle and the second one for a pair of particles or
more precisely for a parity doublet. Our approach is based on wave functions
with well defined tensorial and spinorial properties. Namely, our wave
functions describing an interacting massive particle with higher spin $s$ is
an irreducible skew-symmetric tensor of rank $2n$ with $n=s-{\frac{1}{2}}$
each component of which is a bispinor.

Our approach is simple and straightforward when going from, say, $s={\frac{3%
}{2}}$ to a general half-integer spin $s$, is causal, describes the
anomalous interaction of a particle having spin $s$ and preferred value $g=2$
of the gyromagnetic ratio, has a suitable non-relativistic limit, etc.

The appearance of RWE which consistently describe  pairs of higher spin
particles (parity doublets) instead of single particles might be
advantageous of our approach since most of above mentioned observed
resonances with higher spins have been found to be parity doublets \cite{24}%
. Mathematically, each of these RWE actually define a carrier space of
irreducible representation of the {\it complete} Poincar\'{e} group (i.e.,
the Poincar\'{e} group including discrete transformations $P,T$ and $C $)
which, when considered as a representation space of the {\it proper}
Poincar\'{e} group, corresponds to the carrier space of a reducible
representations isomorphic to a direct sum of two equivalent 
irreducible representations.

We shall  restrict ourselves to massiveinterracting  particles since for massless ones
there are no-go theorems which state that it is impossible to build a consistent theory of 
interaction of such particles with electromagnetic \cite{nogo1} or gravitational \cite{nogo2}
fields in space-time which is
assymptotically flat . However, we present a brief discussion of the massless limit
of free particle equations which appears to be well defined and generates consistent equations
for massless fields with arbitrary spins.  

In Section II we outline  the Rarita-Schwinger theory \cite{25}  for particles
of spin $s={\frac 32}$ and
discuss the troubles with interaction problems. Notice that, contrary to
the statement of paper \cite{27} these troubles cannot be overcome with
using the Singh-Hagen approach \cite{8} (for the proof see Appendix A) .

In Sections III-V we introduce a new formulation of equations for particles
with spin ${\frac 32}$ (which effectively are equations for parity doublets)
which are causal. The massless limit of these equations is discussed in
Section VI. In Section VII we present equations for single particle states,  causality 
aspects of which are discussed in Appendix B .


\section{\bf Rarita-Schwinger equation}

We begin with the most popular formulation of RWE for particle of spin ${%
\frac 32}$ proposed by Rarita and Schwinger \cite{25}. The wave function is a 16-component 
fourvector-bispinor 
$\psi _{(\alpha )}^\mu $ with $\mu =0,1,2,3$ being a four-vector index and $%
\alpha =1,2,3,4$  a bispinor index which will be usually omitted. Then the
RS equation can be written in the form \cite{25} 
$$
\begin{array}{l}
\left( \gamma _\lambda p^\lambda +m\right) \psi ^\mu =0, \\ 
\gamma _\mu \psi ^\mu =0,
\end{array}
\eqno(2.1) 
$$
where $\gamma _\lambda $ are the Dirac matrices acting on the bispinor
index in the following way: $\left( \gamma _\lambda \psi ^\mu \right) _{(\alpha )}=\Sigma
_{\sigma =1}^4\left( \gamma _\lambda \right) _{(\alpha )(\sigma )}\psi
_{(\sigma )}^\mu $.

Contracting the first of equations (2.1) with $\gamma_\mu$ we obtain the compatibility condition
 for the system (2.1): 
$$
p_\mu\psi^\mu=0.\eqno(2.2) 
$$

The RS system (2.1), (2.2) can be rewritten as a single equation 
$$
{\cal F}_\mu =L_{\mu \lambda }\psi ^\lambda =0\eqno(2.3) 
$$
with operator $L_{\mu \nu }$ of the form 
$$
L_{\mu \lambda }= \left( \gamma ^\nu p_\nu +m\right) g_{\mu \lambda }-\gamma
_\mu p_\lambda -\gamma _\lambda p_\mu +\gamma _\mu \left( \gamma ^\nu p_\nu
-m\right) \gamma _\lambda .\eqno(2.4) 
$$

Reducing (2.3) with $\gamma _{\mu }$ and $p_{\mu }$ we get equations
(2.1).

Equation (2.3) admits the Lagrangian formulation. The corresponding Lagrangian $L$
can be written as 
$$
L=\bar{\psi}^{\mu }L_{\mu \nu }\psi ^{\nu },\eqno(2.5) 
$$
where $\bar{\psi}^{\mu }=\psi ^{\mu \dag }\gamma _{0}$.

Let us discuss now the RS equation with interaction. The minimal
interaction with the external e.m. field can be introduced replacing%
\renewcommand{\theequation}{\arabic{section}.\arabic{equation}}%
\setcounter{equation}{7} 
\begin{equation}
p_\mu \rightarrow \pi _\mu =p_\mu -eA_\mu  \label{1}
\end{equation}
in the considered free equation. In order to be sure that this change does not
break the compatibility of our equations we have to make change (\ref{1}) in
the Lagrangian (2.5) whose variation w.r.t. $\bar{\psi}^\mu $ generates the
following equation

\begin{equation}
\begin{array}{l}
\left( \gamma ^\nu \pi _\nu +m\right) \psi ^\mu -\gamma ^\mu \pi _\alpha
\psi ^\alpha -\pi ^\mu \gamma _\alpha \psi ^\alpha 
+\gamma ^\mu \left( \gamma ^\nu \pi _\nu -m\right) \gamma _\lambda \psi
^\lambda =0.
\end{array}
\label{2}
\end{equation}

Contracting (\ref{2}) with $\gamma _\mu $ and $\pi _\mu $ we obtain two
conditions, namely 
\begin{equation}
\gamma _\mu \psi ^\mu =f^\nu \psi _\nu 
\index{3}  \label{3}
\end{equation}
and 
\begin{equation}
\pi _\mu \psi ^\mu =\left( \gamma _\nu \pi ^\nu -{%
\frac 32}m\right) f^\nu \psi _\nu.  \label{4}
\end{equation}
Here 
\[
f^\nu ={\frac{2ie}{3m^2}}\gamma _\mu \tilde{F}^{\mu \nu },\   \tilde{F}%
^{\mu \nu }={\frac 12}\gamma _5\varepsilon ^{\mu \nu \rho \sigma }F_{\rho
\sigma }  
\]
 with $\gamma _5=\gamma _0\gamma _1\gamma _2\gamma _3$ 
and $F^{\nu \sigma }=-{\frac ie}[\pi ^\nu ,\pi ^\sigma ]$ is the strength tensor of
the electromagnetic field.

Using conditions  (\ref{3}), (\ref{4}) equation (\ref{2}) reduces to the form 
\begin{equation}
\left( \gamma _\nu \pi ^\nu +m\right) \psi ^\mu -\left( \pi ^\mu -{\frac m2}%
\gamma ^\mu \right) f^\nu \psi _\nu =0,  \label{5}
\end{equation}
which together with equation (\ref{3}) is equivalent to (\ref{2}).
Equation (\ref{5}) has a non-singular matrix coefficient for the time
derivative and is called the ''true motion equation''.

There are two important physical requirements which have to be imposed to
any RWE for a particle of spin $s$. Namely, that a) the related Cauchy initial
value problem must possess a unique solution depending on 2(2s+1) initial
data functions, and that b) the velocity of propagation of the wave solutions
must not exceed the velocity of light in vacuum.

Condition a) for the  the RS equation is fulfilled due to the following facts. First, 
evolution equation (\ref{5}) is supplemented by constraint (\ref{3}).
One more constraint is generated by equation (\ref{2}) for $\mu =0$: 
\begin{equation}
\pi _a\psi _a+(\gamma _a\pi _a-m)\gamma _b\psi _b=0  \label{6}
\end{equation}
where summation is understood over the repeated indices $a,b=1,2,3$.

Relations (\ref{3}) and (\ref{6}) are compatible with (\ref{5}) and reduce
the 16 components $\psi ^{\mu }$ to  $8$ (i.e., $2(2s+1)$ with $s=3/2$) independent ones.

However, the RS equation does not satisfy requirement b). To show this it is
sufficient to consider the equations (\ref{3})-(\ref{5}) in the eikonal
approximation $\Psi ^\mu =\hat{\psi}^\mu \exp (i\tau n_\nu x^\nu ),\tau
\rightarrow \infty $ where $\hat{\psi}^\mu $ are constants and $n_{\mu \text{
}}$ is a  constant four-vector.  This actually means to substitute the characteristic 
four-vector $n_\mu $ to the
covariant derivatives and keep only leading terms in $n_\mu $.  Then (\ref{5}) reduces 
to a system of linear
homogeneous  algebraic equations. Equating to zero the determinant of matrix defining 
this system we obtain an algebraic equation for $n_\mu .$ Then, if all $n_0$
satisfying this equation are real, the system~(\ref{2}) is hyperbolic and if all $n_0$
satisfy $\frac{n_0^2}{{\bf n}^2}\leq 1$ or $n_0^2-{\bf n}^2\leq 0$ where 
$ {\bf n}^2=n_1^2+n_2^2+n_3^2$, the theory is causal  \cite{15}, \cite{37}.

However it seems that the simplest way to prove acausality of (\ref{3}), (\ref{5}) is to
choose ad hoc special solutions of the form $\Psi ^\mu =p^\mu \phi $ and show that it is acausal. 
In the eikonal approximation such solutions satisfy (\ref{5}) identically provided (%
\ref{3}) is satisfied. On the other hand, equation (\ref{3}) is reduced to
the following form 
\begin{equation}
\left( \gamma _\nu n^\nu -{\frac{2ie}{3m^2}}\gamma _\nu \tilde{F}^{\nu
\sigma }n_\sigma\right) \phi =0.  \label{0005}
\end{equation}

Choosing $n_\mu =(n,0,0,0)$ we conclude that (\ref{0005}) admits non-trivial
solutions for time-like $n_\mu $ which evidently are acausal. Moreover, it is possible to show that
acausal solutions appear even for very small (but non-zero) $F^{\mu \nu }$%
\cite{15}.

Thus the minimally coupled RS equation admits faster-than-light solutions
and  is not in this sense satisfactory. It was shown in \cite{38} that the RS
equation with anomalous interaction is acausal too (for the most recent
analysis of this problem see ref. \cite{38A}).

It is, therefore, still current to search for consistent formulations of RWE for a
particle with spin ${\frac 32}$ and higher. They will be described in Sections
III-VII.

\renewcommand{\theequation}{\arabic{section}.\arabic{equation}} %
\setcounter{equation}{0}

\section{\bf Equations for parity doublets }

The $RS$ equation with spin $\frac{3}{2}$ and its
generalizations have been formulated in terms
of fourvector-bispinor and symmetric tensor-bispinor wave functions
respectively \cite{3}, \cite{8}. 

We shall propose here an approach valid for particles with arbitrary 
 higher half-integer spins $s$ in which the spin-$s$
fermionic field is described by  $\Psi ^{\lbrack \mu _{1}v_{1}][\mu _{2}v_{2}]\ldots \lbrack
\mu _{n}\nu _{n}]}$ - an {\it antisymmetric} irreducible
tensor-spinor of rank $2n$\footnote{%
i.e., the tensor antisymmetric w.r.t. permutations $\mu _{i}$ with $\nu
_{i} $  and symmetric w.r.t. permutations of $[\mu _{i},\nu _{i}]$ with $%
[\mu _{j},\nu _{j}]$ and, moreover, having zero all contractions with $g_{\mu_i
\nu_j }$ and $\varepsilon _{\mu_i \nu_i \mu_j \nu_j }, i,j=1,2,...,n$.}  ($n=s-%
\frac{1 }{2})$ satisfying the condition 
\begin{equation}
\gamma _{\mu_1 }\gamma _{\nu_1 }\Psi ^{\lbrack \lambda \sigma ][\mu
_{1}\nu _{1}]\ldots \lbrack \mu _{n}\nu _{n}]}=0,  \label{3.1}
\end{equation}
where $\gamma _{\lambda }$ and $\gamma _{\sigma }$ are the Dirac matrices.
Moreover, field $\Psi ^{\lbrack \mu _{1}v_{1}][\mu _{2}v_{2}]\ldots }$ is
supposed to satisfy the Dirac equation 
\begin{equation}
(\gamma ^{\lambda }p_{\lambda }-m)\Psi ^{\lbrack \mu _{1}\nu _{1}][\mu
_{2}\nu _{2}]\ldots \lbrack \mu _{n}\nu _{n}]}=0.  \label{3.2}
\end{equation}

A mere consequence of (\ref{3.1}) and (\ref{3.2}) is the following relation 
\begin{equation}
\gamma _{\lambda }\pi _{\sigma }\Psi ^{\lbrack \lambda \sigma ][\mu _{2}\nu
_{2}]\ldots \lbrack \mu _{n}\nu _{n}]}=0.  \label{3.02}
\end{equation}

We shall see that antisymmetric tensors are in many respects more convenient
for constructing RWEs than the usually used symmetric tensors \cite{3}, \cite
{8}, since they more naturally lead to causal equations.

In accordance with its definition, field 
$\Psi ^{[\mu _1\nu _1]\ldots [\mu _n\nu _n]}$ transforms according to the representation 
\begin{equation}
\begin{array}{l}
\lbrack D(s-{\frac 12},0)\oplus D(0,s-{\frac 12})]\otimes [D(\frac 12%
,0)\oplus D(0,\frac 12)] 
\vspace{2mm}
 \\ 
=D(s,0)\oplus D(0,s)\oplus D(s-{\frac 12},{\frac 12})\oplus D(\frac 12,s-{%
\frac 12})
\vspace{2mm}
 \\ \oplus D(s-1,0)\oplus D(0,s-1)
\end{array}
\label{3.5}
\end{equation}
of the Lorentz group, so that it has $16s$ components.

Relation (\ref{3.1}) defines a static constraint, i.e., the constraint which does not involve
derivatives. Expressing $p_0 \Psi^{ [0
\nu_1][\mu_2 \nu_2]\ldots [\mu_n \nu_n]}$ in terms of derivatives w.r.t. the space
variables in (\ref{3.02}) we get the second, dynamical constraint.

Static constraint (\ref{3.1}) suppresses the states corresponding to the
representations $D(s-1,0)$ and $D(0,s-1)$ and relation (\ref{3.02}) reduces
half of the remaining states, so that we have exactly 4(2s+1) independent
components, i.e., twice more than necessary.

Equations (\ref{3.1})- (\ref{3.02}) can be replaced by the following equation 
\begin{equation}
\begin{array}{l}
L^{[\mu _1\nu _1][\mu _2\nu _2]\ldots [\mu _n\nu _n][\lambda _1\sigma
_1][\lambda _2\sigma _2]\ldots [\lambda _n\sigma _n]}\Psi _{[\lambda
_1\sigma _1][\lambda _2\sigma _2]\ldots [\lambda _n\sigma _n]} 
\vspace{2mm}
 \\ 
=(\gamma _\lambda p^\lambda -m)\Psi ^{[\mu _1\nu _1][\mu _2\nu _2]\ldots
[\mu _n\nu _n]}
\vspace{2mm}
 \\ -\frac 1{4s}\sum_\wp (\gamma ^{\mu _1}\gamma ^{\nu _1}-\gamma
^{\nu _1}\gamma ^{\mu _1})p_\lambda \gamma _\sigma \Psi ^{[\lambda \sigma
][\mu _2\nu _2]\ldots [\mu _n\nu _n]}=0
\end{array}
\label{3.8}
\end{equation}
where the symbol $\sum_\wp $ denotes the sum over permutations of subindices 
$(2,...n)$ with $1$ and tensor $\Psi ^{[\mu _1\nu _1][\mu _2\nu _2]\ldots [\mu
_n\nu _n]}$ is supposed to satisfy relation (\ref{3.1}).

Contracting (\ref{3.8}) with $\gamma _\mu \gamma _\nu $ we get an identity
while contraction (\ref{3.8}) with $p_\mu \gamma _\nu $ yields relation (\ref
{3.02}).

It is important to notice that equations (\ref{3.8}) can be derived from a Lagrangian of the form 
\begin{equation}L={\overline{\Psi }}_{[\mu _1\nu _1][\mu _2\nu _2]\ldots [\mu _n\nu
_n]}L^{[\mu _1\nu _1][\mu _2\nu _2]\ldots [\mu _n\nu _n][\lambda _1\sigma
_1][\lambda _2\sigma _2]\ldots [\lambda _n\sigma _n]}\Psi _{[\lambda
_1\sigma _1][\lambda _2\sigma _2]\ldots [\lambda _n\sigma _n]}  \label{3.6}
\end{equation}
with $L^{[\mu _1\nu _1][\mu _2\nu _2]\ldots [\mu _n\nu _n][\lambda _1\sigma
_1][\lambda _2\sigma _2]\ldots [\lambda _n\sigma _n]}\Psi _{[\lambda
_1\sigma _1][\lambda _2\sigma _2]\ldots [\lambda _n\sigma _n]}$ defined in ( 
\ref{3.8}) and the tensor $\Psi ^{\lbrack \lambda \sigma ][\mu
_{1}\nu _{1}]\ldots \lbrack \mu _{n}\nu _{n}]}$ assumed to satisfy (\ref{3.1}).

In the case $s=\frac 32$ this Lagrangian is of the form
\begin{equation}  \label{3.60}
L={\overline{\Psi}}_{[\mu\nu]}L^{[\mu\nu][\lambda
\sigma]}\Psi_{[\lambda\sigma]},
\end{equation}
where 
\begin{equation}  \label{3.71}
\begin{array}{l}
L^{[\mu\nu][\lambda\sigma]}=\frac 12(\gamma^\alpha p_\alpha-m) (g^{\mu
\lambda}g^{\nu\sigma}-g^{\mu \sigma}g^{\nu \lambda})\\ -
 \frac 1{12} (\gamma^\mu\gamma^\nu-\gamma^\nu\gamma^\mu)(p^%
\lambda\gamma^\sigma-p^\sigma\gamma^\lambda).
\end{array}
\end{equation}

We notice that it is always possible to chose such Lagrangian which
generates also simultaneously equation 
(\ref{3.1}) 
 (so that validity of this  equation is not necessary to be  assumed a priori). For $s=3/2$  it
 has the form  
\begin{equation}  \label{3.70}
\begin{array}{l}
L^{[\mu\nu][\lambda\sigma]}=\frac 12(\gamma^\alpha p_\alpha-m) (g^{\mu
\lambda}g^{\nu\sigma}-g^{\mu \sigma}g^{\nu \lambda})
\vspace{2mm}
 \\ + \frac 1{12}%
(p^\mu\gamma^\nu-p^\nu\gamma^\mu)(\gamma^\lambda\gamma^\sigma-\gamma^\sigma%
\gamma^\lambda) - 
\frac 1{12} (\gamma^\mu\gamma^\nu-\gamma^\nu\gamma^\mu)(p^%
\lambda\gamma^\sigma-p^\sigma\gamma^\lambda)
\vspace{2mm}
 \\   
+\frac{1}{24}(\gamma\mu\gamma^\nu-\gamma^\nu\gamma^\mu)\gamma_\rho p^\rho
(\gamma^\lambda \gamma^\sigma- \gamma^\sigma \gamma^\lambda).
\end{array}
\end{equation}

The corresponding propagator is given by 
\begin{equation}
\begin{array}{l}
G^{[\mu \nu ][\lambda \sigma ]}=\frac{\gamma ^\alpha p_\alpha +m}{p_\lambda
p^\lambda -m^2}[(g^{\mu \lambda }g^{\nu \sigma }-g^{\mu \sigma }g^{\nu
\lambda })+\frac 16(p^\mu \gamma ^\nu -p^\nu \gamma ^\mu )(\gamma ^\lambda
\gamma ^\sigma -\gamma ^\sigma \gamma ^\lambda )
\vspace{2mm}
 \\ 
-\frac 16(\gamma ^\mu \gamma
^\nu -\gamma ^\nu \gamma ^\mu )(p^\lambda \gamma ^\sigma -p^\sigma \gamma
^\lambda )  
+\frac 1{12}(\gamma \mu \gamma ^\nu -\gamma ^\nu \gamma ^\mu )\gamma _\rho
p^\rho (\gamma ^\lambda \gamma ^\sigma -\gamma ^\sigma \gamma ^\lambda )].
\end{array}
\label{3.700}
\end{equation}

Let us comment that from the representation point of view our equations (\ref
{3.8}) are equivalent to those proposed by Lomont and Moses \cite{31} (see
also \cite{32} and \cite{33}).
 However, due to their different forms they 
essentially differ  in the interaction context. Whereas our tensor-spinorial formulation 
(\ref{3.8}) seems to be suitable and very convenient for systematic and consistent introduction 
of various types of interactions,
the Lomont-Moses formulation is consistent for description of free particles only.
\renewcommand{\theequation}{\arabic{section}.\arabic{equation}} %
\setcounter{equation}{0}

\section{Minimal and anomalous interactions}

The minimal interaction with an external electromagnetic field can be
introduced by using replacement (\ref{1}) in the Euler--Lagrange equation (%
\ref{3.8}). As a result we obtain 
\begin{equation}
\begin{array}{l}
(\gamma _\lambda \pi ^\lambda -m)\Psi ^{[\mu _1\nu _1][\mu _2\nu _2]\ldots
[\mu _n\nu _n]}
\vspace{2mm}
 \\ 
-\frac 1{4s}\sum_\wp (\gamma ^{\mu _1}\gamma ^{\nu _1}-\gamma ^{\nu
_1}\gamma ^{\mu _1})\pi _\lambda \gamma _\sigma \Psi ^{[\lambda \sigma ][\mu
_2\nu _2]\ldots [\mu _n\nu _n]}=0.
\end{array}
\label{3.11}
\end{equation}

Contracting (\ref{3.11}) with $\pi _\mu \gamma _\nu $ and using (\ref{3.1})
we obtain the following relation 
\begin{equation}
\begin{array}{l}
\pi _\lambda \gamma _\sigma \Psi ^{[\lambda \sigma ][\mu _1\nu _1]\ldots
[\mu _{n-1}\nu _{n-1}]}
\vspace{2mm}
 \\ 
=\frac{ie}{2m}\left(F_{\lambda \sigma }- \gamma^\nu\gamma_\lambda F{_{\sigma
\nu }}\right)\Psi ^{[\lambda \sigma ][\mu _1\nu _1]\ldots [\mu _{n-1}\nu
_{n-1}]}.
\end{array}
\label{3.12}
\end{equation}

In view of (\ref{3.1}) and (\ref{3.12}) equation (\ref{3.11}) can be written as 
\begin{equation}
\begin{array}{l}
\left( \gamma _\mu \pi ^\mu -m\right) \Psi ^{[\mu _1\nu _1][\mu _2\nu
_2]\ldots [\mu _n\nu _n]} 
\vspace{2mm}
 \\ 
=\frac{ie}{4sm}\sum_\wp \left( \gamma ^{\mu _1}\gamma ^{\nu _1}-\gamma ^{\nu
_1}\gamma ^{\mu _1}\right) (F_{\lambda \sigma }  
-\gamma ^\lambda \gamma _\sigma F_{\alpha \lambda })\Psi ^{[\sigma \alpha
][\mu _1\nu _1]\ldots [\mu _{n-1}\nu _{n-1}]}.
\end{array}
\label{3.131}
\end{equation}
Equations (\ref{3.11}), (\ref{3.1}) and (\ref{3.131}) are suitable for  description of a
 particle with arbitrary half-integer spin $s$. We
shall discuss these equations in detail for the simplest case $s={\frac 32}$%
. However, the obtained results remain true for arbitrary $s$.

For $s={\frac 32}$ the corresponding tensor-spinor function 
has only one pair of indices and thus equations (%
\ref{3.131}), (\ref{3.1}) are reduced to the following form 
\begin{equation}
\begin{array}{l}
{\cal F}^{\mu \nu }=\left( \gamma _\mu \pi ^\mu -m\right) \Psi ^{[\mu \nu ]}

\vspace{2mm}
 \\ 
-\frac{ie}{6m}\left( \gamma ^\mu \gamma ^\nu -\gamma ^\nu \gamma ^\mu
\right) \left( F_{\lambda \sigma }\Psi ^{[\lambda \sigma ]}+\gamma ^\lambda
\gamma _\sigma F_{\alpha \lambda }\Psi ^{[\sigma \alpha ]}\right) 
\end{array}
\label{3.13}
\end{equation}
and
\begin{equation}
\gamma _\mu \gamma _\nu \Psi ^{[\mu \nu] }=0.  \label{3.133}
\end{equation}

Equation (\ref{3.13}) is equivalent to the system 
\begin{equation}
\begin{array}{l}
{\cal F}_{+}^{\mu \nu }=\gamma _\lambda \pi ^\lambda \Psi _{+}^{[\mu \nu
]}-m\Psi _{-}^{[\mu \nu ]}=0,
\end{array}
\label{3.134}
\end{equation}
and
\begin{equation}
\begin{array}{l}
{\cal F}_{-}^{\mu \nu }=\gamma _\lambda \pi ^\lambda \Psi _{-}^{[\mu \nu
]}-m\Psi _{+}^{[\mu \nu ]}
+\frac 16\left( \gamma ^\mu \gamma ^\nu -\gamma ^\nu \gamma ^\mu \right)
\gamma _\lambda \pi _\sigma \Psi _{-}^{[\lambda \sigma ]}=0,
\end{array}
\label{3.135}
\end{equation}
where ${\cal F}_{\pm }^{\mu \nu }={\cal F}^{\mu \nu }\pm \frac 12\gamma _5{%
\varepsilon ^{\mu \nu }}_{\rho \sigma }{\cal F}^{\rho \sigma },\,\Psi _{\pm
}^{[\mu \nu ]}={\Psi }^{[\mu \nu ]}\pm \frac 12\gamma _5{\varepsilon ^{\mu
\nu }}_{\rho \sigma }F^{[\rho \sigma ]}$.

Solving (\ref{3.134}) for $\Psi _{-}^{[\mu \nu ]}$ and using (\ref{3.135})
we obtain the following relation 
\begin{equation}
\begin{array}{l}
\left( \pi _\lambda \pi ^\lambda -\frac{ie}2\gamma _\lambda \gamma _\sigma
F^{\lambda \sigma }-m^2\right) \Psi _{+}^{[\mu \nu ]}  
-\frac i6\left( \gamma ^\mu \gamma ^\nu -\gamma ^\nu \gamma ^\mu \right)
F_{\lambda \sigma }\Psi _{+}^{[\lambda \sigma ]}=0.
\end{array}
\label{3.17}
\end{equation}

Formula (\ref{3.17}) presents a nice second-order hyperbolic differential
equation whose solutions $\Psi _{+}^{[\mu \nu ]}$ are causal. The same is
true for components $\Psi _{-}^{[\mu \nu ]}$ , expressed in terms of 
$\Psi _{+}^{[\mu \nu ]}$ via relation (\ref{3.135}), as well as for $\Psi
^{[\mu \nu ]}$ which is the sum of $\Psi _{+}^{[\mu \nu ]}$ and $\Psi
_{-}^{[\mu \nu ]}$.

Let us remark that equation (\ref{3.17}) can be expressed in the form 
\begin{equation}
(\pi _\mu \pi ^\mu -m^2-\frac{ige}2S^{\mu \nu }F_{\mu \nu })\Psi
_{+}^{[\lambda \sigma ]}=0,  \label{3.19}
\end{equation}
where $g={\frac 23}$, i.e., is reciprocal to $s$, and $S_{\mu \nu }$ are spin generators 
of the Lorentz
group which act on the tensor-bispinor $\Psi _{+}^{[\lambda \sigma ]}$ in the following way:

\begin{equation}
\begin{array}{l}
S^{\rho \sigma }\Psi _{+}^{[\mu \nu ]}=\frac i4\left[ \gamma ^\rho ,\gamma
^\sigma \right] \Psi _{+}^{[\mu \nu ]} 
\vspace{2mm}
 \\  
+i\left( g^{\rho \mu }\Psi _{+}^{[\delta \nu ]}-g^{\delta \mu }\Psi
_{+}^{[\rho \nu ]}-g^{\rho \nu }\Psi _{+}^{[\sigma \mu ]}+g^{\sigma \nu
}\Psi _{+}^{[\rho \mu ]}\right).
\end{array}
\label{03}
\end{equation}
Formula (\ref{3.19}) generalizes the Zaitsev-Feynman-Gell-Mann equation for
electron \cite{23} to the case of particles with spin $\frac 32$. It
describes a charged particle whose gyromagnetic ratio $g$ is $\frac 1s=\frac %
23$.

Following Pauli \cite{39} we can generalize equation (\ref{3.8}) to that
with ''anomalous'' interaction by adding to it a term $L^{[\mu \nu ][\rho \sigma
]}(F)$ linear in $F^{\mu \nu }$, i.e., by changing 
\[
L^{[\mu \nu ][\rho \sigma ]}\rightarrow L^{[\mu \nu ][\rho \sigma ]}(\pi
)+L^{[\mu \nu ][\rho \sigma ]}(F). 
\]

This term can be found as a linear combination of  all
antisymmetric tensors linear in $F^{\mu \nu }$. The complete set of such tensors
 can be derived in terms of tensors  
$F^{[\mu \nu ]}$, $\varepsilon ^{\mu \nu \rho \sigma }$, $%
g_{\mu \nu }$ and vectors $\gamma ^\mu $ 
 and is given by: 
\begin{equation}
\begin{array}{l}
L_1^{[\mu \nu ][\rho \sigma ]}=\gamma _\lambda \gamma _\alpha F^{\lambda
\alpha }(g^{\mu \rho }g^{\nu \sigma }-g^{\mu \sigma }g^{\nu \rho }),
\vspace{2mm}
 \\ 
L_2^{[\mu \nu ][\rho \sigma ]}=i\gamma _5\gamma _\lambda \gamma _\alpha
F^{\lambda \alpha }\varepsilon ^{\mu \nu \rho \sigma }, 

 \vspace{2mm}
 \\  
L_3^{[\mu \nu ][\rho \sigma ]}=F^{\mu \rho }g^{\nu \sigma }-F^{\nu \rho
}g^{\mu \sigma }-F^{\mu \sigma }g^{[\nu \rho ]}+F^{\nu \sigma }g^{\mu \rho },

\vspace{2mm}
 \\  
L_4^{[\mu \nu ][\rho \sigma ]}=i\gamma _5(F^{\alpha \mu }{\varepsilon
_\alpha }^{\nu \rho \sigma }-F^{\alpha \nu }{\varepsilon _\alpha }^{\mu \rho
\sigma } 
+F^{\alpha \rho }{\varepsilon _\alpha }^{\sigma \mu \nu }-F^{\alpha \sigma }{%
\varepsilon _\alpha }^{\rho \mu \nu }), 
\vspace{2mm}
 \\ 
L_5^{[\mu \nu ][\rho \sigma ]}=F^{\nu \rho }\gamma ^\mu \gamma ^\sigma
-F^{\mu \rho }\gamma ^\nu \gamma ^\sigma  
-F^{\nu \sigma }\gamma ^\mu \gamma ^\rho +F^{\mu \sigma }\gamma ^\nu \gamma
^\rho , 
\vspace{2mm}
 \\ 
L_6^{[\mu \nu ][\rho \sigma ]}=\gamma ^\mu \gamma _\lambda F^{\rho \lambda
}g^{\rho \sigma }-\gamma ^\nu \gamma _\lambda F^{\rho \lambda }g^{\mu \sigma
}  
-\gamma ^\mu \gamma _\lambda F^{\sigma \lambda }g^{\nu \rho }+\gamma ^\nu
\gamma _\lambda F^{\sigma \lambda }g^{\mu \rho }, 
\vspace{2mm}
 \\ 
L_7^{[\mu \nu ][\rho \sigma ]}=F^{\mu \nu }(\gamma ^\rho \gamma ^\sigma
-\gamma ^\sigma \gamma ^\rho )+F^{\rho \sigma }(\gamma ^\mu \gamma ^\nu
-\gamma ^\nu \gamma ^\mu ).
\end{array}
\label{3.23}
\end{equation}
Hence the general form of $L^{[\mu \nu ][\rho \sigma ]}(F)$ can be written as 
\begin{equation}
L^{[\mu \nu ][\rho \sigma ]}(F)=\sum_{n=1}^7\alpha _nL_n^{[\mu \nu ][\rho
\sigma ]},  \label{3.24}
\end{equation}
where $\alpha _n$ are arbitrary constants.

A natural condition which can be imposed on $L^{[\mu \nu ][\rho \sigma ]}(F)$
is that the equation with anomalous interaction should be compatible with
relations (\ref{3.1}) which suppress spin $\frac{1}{2}$ states. The {\it %
sufficient} conditions which guarantee this property of $L^{[\mu \nu ][\rho
\sigma ]}$ are 
\begin{equation}
\begin{array}{l}
\gamma _{\mu }\gamma _{\nu }L^{[\mu \nu ][\rho \sigma ]}(F)=0
\end{array}
\label{3.25}
\end{equation}
and
\begin{equation}\begin{array}{l}\pi_{\mu }
\gamma _{\nu }L^{[\mu \nu ][\rho \sigma ]}(F)=0.\end{array}\label{nid}\end{equation}
Substituting expression (\ref{3.24}) into the conditions (\ref{3.25}) and (\ref{nid}) we obtain 
\[
4\alpha _1=4\alpha _2=\alpha _3=\alpha _4=\frac{2k}3,\quad \alpha _5=\alpha
_6=\alpha _7=0 ,
\]
where $k$ is so far  an arbitrary parameter.
Consequently the equation with anomalous interaction is of the form 
\begin{equation}
\begin{array}{l}
(\gamma ^\lambda \pi _\lambda -m)\Psi ^{[\mu \nu ]}-\frac 16(\gamma ^\mu
\gamma ^\nu -\gamma ^\nu \gamma ^\mu )\pi _\alpha \gamma _\sigma \Psi
^{[\alpha \sigma ]} 
\vspace{2mm}
 \\ 
+\frac{iek}{3m}(\frac 14\gamma _\alpha \gamma _\sigma F^{\alpha \sigma }{%
\Psi }_{+}^{[\mu \nu ]}+{F_\alpha }^\mu {\Psi }_{+}^{[\nu \alpha ]}-{%
F_\alpha }^\nu \hat{\Psi}_{+}^{[\mu \alpha ]}).
\end{array}
\label{3.26}
\end{equation}

Contracting (\ref{3.26}) with $\gamma _\mu \gamma _\nu $ and $\pi _\mu
\gamma _\nu $ we get again conditions (\ref{3.1}) and (\ref{3.12}), which
enable us to write equation (\ref{3.26}) as a system which consists of  (\ref{3.134}) and
the equation 
\begin{equation}
\begin{array}{l}
\gamma _\lambda \pi ^\lambda \Psi _{-}^{[\mu \nu ]}-m\Psi _{+}{}^{[\mu \nu
]}-\frac 16\left( \gamma ^\mu \gamma ^\nu -\gamma ^\nu \gamma ^\mu \right)
\gamma _\lambda \pi _\sigma \Psi _{-}^{[\lambda \sigma ]} 
\vspace{2mm}
 \\ 
+\frac{iek}{3m}(\frac 14\gamma _\alpha \gamma _\sigma F^{\alpha \sigma }{%
\Psi }_{+}^{[\mu \nu ]}+{F_\alpha }^\mu {\Psi }_{+}^{[\nu \alpha ]}-{%
F_\alpha }^\nu \hat{\Psi}_{+}^{[\mu \alpha ]})=0.
\end{array}
\label{3.27}
\end{equation}
Solving  equation (\ref{3.134}) for $\Psi _{-}^{[\mu \nu ]}$ and using (\ref{3.27}) we obtain
the second order equation (\ref{3.19}) in which, however, $g=\frac 23(1+k).
$

Thus the anomalous interaction  causes only one thing, namely, that the gyromagnetic ratio
 $g$ in (\ref{3.17}) which in minimal interaction case was fixed and equal to $\frac 1s$ becomes 
 arbitrary, but the form of the equation remains the same. The possibility  of changing $g$ 
without changing the form of the equation 
seems to be an attractive
feature of the proposed approach. 

We recall that even in the case of the Dirac equation introduction of the
anomalous interaction leads to a very essential complication of the theory.
Indeed, the Dirac equation with minimal interaction is mathematically
equivalent to Zaitsev-Feynman-Gell-Mann equation, the explicit form of which can be
obtained from (\ref{3.19}) by changing ${\Psi }_{+}^{[\mu \nu ]}\rightarrow
\psi ,\,g\rightarrow 2,\,S^{\mu \nu }\rightarrow \frac 14\left[ \sigma ^\mu
,\sigma ^\nu \right] $, where $\psi $ is a two-component spinor and $\sigma
^\mu $ are the Pauli matrices. In the case of anomalous interaction the
related second-order equation (i.e., the analog of (\ref{3.19})) includes a
second order polynomial in $F^{\mu \nu }$ and derivatives of $F^{\mu
\nu }$ w.r.t. $x_\lambda $ as well, which does not happen in our approach.

Taking $k=2$ we can get the
gyromagnetic ratio $g$ equal to $2$, i.e., to its  "natural value" (see, e.g.,  \cite{27}).

\renewcommand{\theequation}{\arabic{section}.\arabic{equation}} %
\setcounter{equation}{0}

\section{Foldy-Wouthuysen reduction}

In order to analyze a non-relativistic approximation of equation (\ref{3.26})
it is convenient to make the Foldy-Wouthuysen reduction and express the
corresponding Hamiltonian in a power series of $\frac 1m$. For this purpose
we shall introduce the following notations
\begin{equation}
\begin{array}{l}
\Psi =\text{column}\ (\Psi ^{23},\Psi ^{31},\Psi ^{12},\Psi ^{01},\Psi ^{02},\Psi
^{03}), 
\vspace{2mm}
 \\ 
\tilde{S}_{\mu \nu }=I_4\otimes S_{\mu \nu },\ \hat{S}^{\mu \nu }=\tilde{S}%
_{\mu \nu }+\frac i4[\hat{\gamma}^\mu ,\hat{\gamma}^\nu ], 
\vspace{2mm}
 \\ 
\hat{\gamma}_\mu =\gamma _\mu \otimes I_6,\ \ \hat{\sigma}_2=\left( 
\begin{array}{cc}
0 & -iI_{12} 
 \\ 
iI_{12} & 0
\end{array}
\right) , 
\vspace{2mm}
 \\  
S_{ab}=\varepsilon _{abc}\left( 
\begin{array}{cc}
s_c & 0 \\ 
0 & s_c
\end{array}
\right) ,\quad S_{0a}=\left( 
\begin{array}{cc}
0 & -s_a \\ 
s_a & 0
\end{array}
\right) ,
\end{array}
\label{3.28}
\end{equation}
where $I_{12},I_6$, and $I_4$ are the $12\times 12$, $6\times 6$ and $%
4\times 4$ unit matrices respectively, and $s_c$ are $3\times 3$ matrices
elements of which are $(s_c)^{ab}=i\varepsilon _{abc}$.

Then equation (\ref{3.27}) multiplied by $\tilde{\gamma}_0$ reads 
\begin{equation}
i\frac \partial {\partial t}\Psi =H\Psi ,  \label{3.30}
\end{equation}
where 
\begin{equation}
\begin{array}{l}
H=\hat{\gamma}_0\hat{\gamma}_a\pi _a+\hat{\gamma}_0m+eA_0 
+\hat{\gamma}_0(1+i\hat{\gamma}_5\hat{\sigma}_2)\frac e{4m}(g\hat{S}_{\mu
\nu }-i\hat{\gamma}_\mu \hat{\gamma}_\nu )F^{\mu \nu }.
\end{array}
\label{3.31}
\end{equation}

To simplify calculations we suppose that $\frac{\partial F^{ab}}{\partial x_c}<<1\ ,a,b,c=1,2,3$, and $g=2$. Then, transforming $H\rightarrow H^{\prime
}=VHV^{-1}+iV\frac \partial {\partial t}V^{-1}$ where $V=\exp
(iS_3)\exp (iS_2)\exp (iS_1)$ with
\[
\begin{array}{l}
S_1=-\frac im\hat{\gamma}_0\left( 1+i\hat{\sigma}_2\hat{\gamma}_5)\right) 
\hat{\gamma}^a\pi ^a, 
\vspace{2mm}
 \\ 
S_2=-\frac{\gamma _5}{4m^2}\left( \pi ^2-e\hat{S}_{\mu \nu }F^{\mu
\nu }\right)  
-i\frac{\gamma _0\gamma _5}{8m^3}\left[ e(p_aE_a+E_ap_a)-2i\frac{\partial 
\hat{S}_{\mu \nu }F^{\mu \nu }}{\partial t}\right] ,
\vspace{2mm}
 \\ 
S_3=\frac{ig}2\gamma _0\varepsilon _{abc}\hat{S}^{ab}\pi ^c
\end{array}
\]
and omitting terms of the order of $\frac 1{m^3}$ we finally obtain\footnote{%
The only  term in (\ref{3.32}) which is of order $\frac 1{m^3}$, i.e., the term ${\frac{\pi ^4}{8m^3}}$,
should be present in as much as it is of 
the same order $\frac 1{c^2}$ as three the last terms (c is the speed of light).  
Using the Heaviside units in wich h=c=1 leads to implicite dependence of the Hamiltonian on $c$.}   
\begin{equation}
\begin{array}{l}
H^{\prime }=\hat{\gamma}_0\left( m+\frac{\pi ^2}{2m}-{\frac{\pi ^4}{8m^3}}-%
\frac em\vec{S}\cdot \vec{H}\right) +eA_0+\frac e{2m^2}\vec{S}\cdot (\vec{%
\pi}\times \vec{E}-\vec{E}\times \vec{\pi})
\vspace{2mm}
 \\ -\frac e{12m^2}Q^{ab}\frac{%
\partial E_a}{\partial x_b}-\frac{es(s+1)}{6m^2}\vec{\nabla}\cdot {\vec{E}}.
\end{array}
\label{3.32}
\end{equation}
Here $\vec{S}$ denotes a vector $(S_1,S_2,S_3)$ with $S_a=\frac 12\varepsilon
_{abc}\hat{S}_{bc}$, $Q_{ab}=3\left[ S_a,S_b\right] _{+}-2s(s+1)\delta
_{ab}\ (s={\frac 32})$ is the quadrupole interaction tensor, and $E_a$ and $%
H_a$ denote components of the electric and magnetic fields vectors. 

All terms of Hamiltonian (\ref{3.32}) have a clear physical meaning. 
For positive  energy solutions they have the following interpretation:
 $m+{%
\frac{\pi ^{2}}{2m}}+eA_{0}$ represents the Schr\"{o}dinger Hamiltonian
 with the rest energy term,  $-{\frac{\pi ^{4}}{8m^{3}}}$ the relativistic correction to the kinetic
energy, $\frac{e}{m}\vec{S}%
\cdot \vec{H}$ is the Pauli coupling , - $\frac{e}{2m^{2}}\vec{S}\cdot (%
\vec{\pi}\times \vec{E}-\vec{E}\times \vec{\pi})$ is the spin-orbit coupling, $-%
\frac{e}{12m^{2}}Q^{ab}\frac{\partial E_{a}}{\partial x_{b}}$ is the quadrupole coupling and
-$\frac{e(s+1)}{6m^{2}}\vec{\nabla}\cdot {\vec{E}}$ is the Darwin coupling.

Let us  remark that all equations starting with (\ref{3.131}) up to
(\ref{3.32}) can easily be extended to the case with {\it arbitrary} half-integer spin $s$. 
As a result we obtain the quasirelativistic Hamiltonian (%
\ref{3.32}) which is of the same form but with $\vec{S}$ corresponding to appropriate 
spin matrices for the considered spin $s$.

\renewcommand{\theequation}{\arabic{section}.\arabic{equation}} %
\setcounter{equation}{0}

\section{The massless  case}

It is well known that relativistic wave equations for massless particles
with higher spins cannot be generally obtained from those for massive 
particles by taking  the limit $m\rightarrow 0$
 \cite{4}. Here we demonstrate that
tensor-spinorial equations (\ref{3.1}) and (\ref{3.8}) have similiar properties like the Dirac equation, 
namely, that they have a clear physical
meaning for $m=0$ provided some additional constraints are imposed on their solutions. 

We begin with spin $s=\frac 32$. Taking equation (\ref{3.8}) appropriate for this case, setting in it $m=0$   and 
supposing that  the condition  
\begin{equation}
\gamma _\nu \Psi ^{[\mu \nu ]}=0  \label{002}
\end{equation}
is true, we obtain the equation
\begin{equation}
\gamma ^\alpha p_\alpha \Psi ^{[\mu \nu ]}=0  \label{01}
\end{equation}
which describes a
massless field whose helicities are $\pm \frac 32$ and energy signes are $%
\pm 1.$ This can be shown in the following way.

Reducing (\ref{01}) with $\gamma _\nu $ and using (\ref{002}) we
get the condition 
\begin{equation}
p_\nu \Psi ^{[\mu \nu ]}=0.  \label{07}
\end{equation}

It follows from (\ref{002}) and (\ref{07}) that 
\begin{equation}
\varepsilon _{\mu \nu \rho \sigma }\gamma ^\nu \Psi ^{[\rho \sigma ]}=0
\label{04}
\end{equation}
and 
\begin{equation}
\varepsilon _{\mu \nu \rho \sigma }p^\nu \Psi ^{[\rho \sigma ]}=0.
\label{05}
\end{equation}
In other words field $\Psi ^{[\mu \nu ]}$ satisfies both the massless Dirac
equation (\ref{01}) and the Maxwell equations (\ref{07}) and (\ref{05}).

Conditions (\ref{04}) reduces the number of independent components of $\Psi
^{\mu \nu }$ to $8$ while relations (\ref{07}) reduce this number to $4$. To
prove that solutions of (\ref{01}), (\ref{002}) correspond to helicities $%
\pm \frac 32$ relations (\ref{03}) and (\ref{05}) should be used from which follow
that 
\begin{equation}
\varepsilon _{abc}S^{ab}p_c\Psi ^{[\mu \nu ]}=\frac 34i\varepsilon
_{abc}\gamma ^a\gamma ^bp_c\Psi ^{[\mu \nu ]}\equiv \frac 32i\gamma _5\gamma
_0\gamma ^ap_a\Psi ^{[\mu \nu ]}.  \label{09}
\end{equation}

In accordance with (\ref{01}) and (\ref{09}) the eigenvalues of the related
helicity operator coincide with eigenvalues of the energy sign operator
multiplied by $\pm \frac 32.$ Thus solutions of equations (\ref{01}) and (\ref
{002}) belong to the carrier space of the irreducible representation $D^{+}(%
\frac 32)\oplus D^{-}(\frac 32)\oplus D^{+}(-\frac 32)\oplus D^{-}(-\frac 32%
)$ of the Poincar\'{e} group, where $D^\epsilon (\lambda )$ denotes
representation corresponding to energy sign $\epsilon $ and to helicity $%
\lambda .$ Imposing the additional constraints $(1+i\gamma _5)\Psi ^{[\mu
\nu ]}=0$ or $(1-i\gamma _5)\Psi ^{[\mu \nu ]}=0$ it is possible to reduce
this representation to $D^{+}(\frac 32)\oplus D^{-}(-\frac 32)$ or $D^{-}(%
\frac 32)\oplus D^{+}(-\frac 32)$. In other words, relations (\ref{01}) and (%
\ref{002}) form a natural generalization of the massless Dirac equation
to the case of spin $\frac 32$.

We note that the ansatz 
\begin{equation}
\Psi ^{[\mu \nu ]}=p^\mu \Psi ^\nu -p^\nu \Psi ^{\mu \text{ }} , \label{010}
\end{equation}
where $\Psi ^{\mu \text{ }}$ is a vector-spinor satisfying the condition $%
\gamma _\lambda \Psi ^{\lambda \text{ }}=0$ reduces (\ref{002}), (\ref{01})
to the massless RS equation for $\Psi ^{\mu \text{ }}:$%
\[
\gamma ^\alpha p_\alpha \Psi ^\mu =0,\qquad \gamma _\lambda \Psi ^{\lambda 
\text{ }}=0.
\]

Equation (\ref{010}) is invariant w.r.t. the gauge transformation $\Psi
^{\lambda \text{ }}\rightarrow \Psi ^{\lambda \text{ }}+\frac{\partial
\varphi }{\partial x_\lambda }$, where $\varphi $ is an arbitrary solution of
the massless Dirac equation $\gamma ^\alpha p_\alpha \varphi =0.$

Analogously, starting with equations (\ref{3.1}), (\ref{3.8}) for
arbitrary spin we come to the following equations for the 
massless field with spin $s=\frac{2n+1}2$ 
\[\ba{l}
\gamma ^\alpha p_\alpha \Psi ^{[\mu _1\nu _1][\mu _2\nu _2]...[\mu _n\nu
_n]}=0,
\vspace{2mm}
\\ \gamma _\alpha \Psi ^{[\alpha \nu _1][\mu _2\nu _2]...[\mu _n\nu
_n]}=0. \ea
\]
Like solutions of (\ref{01}) and (\ref{002}), the related wave function $\Psi ^{[\mu
_1\nu _1][\mu _2\nu _2]...[\mu _n\nu _n]}$ has only four independent
components corresponding to states with helicities $\pm s$ and energy
signs $\pm 1.$\renewcommand{\theequation}{\arabic{section}.\arabic{equation}}
\setcounter{equation}{0}

\section{Single particle Equations}

As was shown, equations (\ref{3.1}) and (\ref{3.2}) describe a doublet of
relativistic particles with spin $s$. In order to find the Poincar\'{e} and parity 
invariant equation for a single
particle it is necessary to impose on $\Psi ^{\lbrack \mu _{1}\nu _{1}][\mu
_{2}\nu _{2}]...[\mu _{n}\nu _{n}]}$ an additional condition which annuls
half of the physical components. It can be taken in the form
\begin{equation}
p_\mu \Psi ^{[\mu \nu _1][\mu _2\nu _2]...[\mu _n\nu _n]}=0.  \label{4.0}
\end{equation}
The resulting system, i.e., (\ref{3.1}), (\ref{3.2}) and (\ref{4.0}) obviously
satisfies all required invariance properties and describes a particle of
arbitrary half-integer spin $s=\frac{2n+1}2$. 

In the case $s={\frac %
32}$ this system is reduced to the equations 
\begin{equation}
(\gamma ^\lambda p_\lambda -m)\Psi ^{[\mu \nu ]}=0,  \label{4.1}
\end{equation}
\begin{equation}
\gamma _\mu \gamma _\nu \Psi ^{[\mu \nu ]}=0,  \label{4.2}
\end{equation}
\begin{equation}
p_\mu \Psi ^{[\mu \nu ]}=0.  \label{4.3}
\end{equation}

In the rest frame $p=(m,0,0,0)$ relation (\ref{4.3}) reduces to $%
m\Psi ^{[oa]}=0,$ which implies $\Psi ^{[oa]}=0$. Thus (\ref{4.3}) annuls half 
 of the components of the wave function. On the other hand, in the
rest frame condition (\ref{4.2}) can be written as 
\begin{equation}
\vec{S}^2\Psi =s(s+1)\Psi ,\quad s=\frac 32,  \label{4.4}
\end{equation}
where $\Psi =column(\Psi ^{[23]},\Psi ^{[31]},\Psi ^{[12]})$ and $\vec{S}%
=\left( S_{23},S_{31},S_{12}\right) $ is the total spin for the
tensor-spinor wave function, components of which are given in (\ref{03}).

The system of equations (\ref{4.1})-(\ref{4.3}) can be replaced by one equivalent 
equation which is of the form
\begin{equation}
\begin{array}{l}
(\gamma ^\lambda p_\lambda -m)\Psi ^{[\mu \nu ]}+\gamma ^\nu p_\lambda \Psi
^{[\lambda \mu ]}-\gamma ^\mu p_\lambda \Psi ^{[\lambda \nu ]}-\frac 12%
[\gamma ^\mu p^\nu
\vspace{2mm}
 \\ 
-\gamma ^\nu p^\mu -(\gamma ^\mu \gamma ^\nu -\gamma ^\nu \gamma ^\mu
)\left( \frac 12\gamma _\lambda p^\lambda -\frac m3\right) ]\gamma _\lambda
\gamma _\sigma \Psi ^{[\lambda \sigma ]}=0.
\end{array}
\label{4.5}
\end{equation}

Indeed, reducing (\ref{4.5}) with $\gamma _\mu \gamma _\nu $ and $p_\mu \gamma _\nu $
we get the system (\ref{4.1})-(\ref{4.3}). On the other hand, reducing (%
\ref{4.5}) with $\gamma _\nu $ and denoting $\gamma _\nu \Psi ^{[\mu \nu
]}$ by $\Psi ^\mu $ we obtain the $RS$ equation (2.3) as an
algebraic consequence of (\ref{4.5}). However, equation (\ref{4.5}) is not
of the Euler-Lagrange type.

In order to find a Lagrangian generating equations (\ref{4.1})-(\ref{4.3}) one should
add an auxiliary field. Using this old idea of Fierz and Pauli \cite{3}  the desired Lagrangian is given by
\begin{equation}
L=L^{TS}+L^{RS}+L^{CR} , \label{4.6}
\end{equation}
where $L^{TS}$ is the Lagrangian of the tensor-spinor field defined in (\ref{3.60}), (%
\ref{3.70}), $L^{RS}$ is the Rarita-Schwinger Lagrangian given in (2.5) and $L^{CR}$
is the "crossed interaction'' Lagrangian of the form 
\begin{equation}
\begin{array}{l}
L^{CR}=-\bar{\Psi}_{[\mu \nu ]}p^\mu \Psi ^\nu +\bar{\Psi}_\mu p_\nu \Psi
^{[\mu \nu ]} 
\vspace{2mm}
 \\ 
-\frac 1{12}(\bar{\Psi}_{[\mu \nu ]}\gamma ^\mu \gamma ^\nu (p_\lambda
-\gamma _sp^s\gamma _\lambda )\Psi ^\lambda 
-\bar{\Psi}^\lambda (p_\lambda -\gamma _sp^s\gamma _\lambda )\gamma ^\mu
\gamma ^\nu \Psi _{[\mu \nu ]}).
\end{array}
\label{4.7}
\end{equation}

Variation of Lagrangian (\ref{4.6}) w.r.t. $\tilde{\Psi}_{[\mu \nu ]}$ and $%
\tilde{\Psi}_\mu $ yields  two equations, namely 
\begin{equation}
\begin{array}{l}
(\gamma _\lambda p^\lambda -m)\Psi ^{[\mu \nu ]}-p^\mu \Psi ^\nu +p^\nu \Psi
^\mu  
+\frac 1{12}(\gamma ^\mu \gamma ^\nu -\gamma ^\nu \gamma ^\mu )(f-2p_\lambda
\gamma _\sigma \Psi ^{[\lambda \sigma ]})
\vspace{2mm}
 \\ 
+\frac 1{24}(\gamma ^\mu \gamma ^\nu -\gamma ^\nu \gamma ^\mu )(\gamma
_\lambda p^\lambda -m)\gamma _\lambda \gamma _\sigma \Psi ^{[\lambda \sigma
]}=0
\end{array}
\label{4.8}
\end{equation}
and 
\begin{equation}
\begin{array}{l}
(\gamma _\lambda p^\lambda +m)\Psi ^\mu -\gamma ^\mu (f+m\gamma _\lambda
\Psi ^\lambda )-p^\mu \gamma _\lambda \Psi ^\lambda 
\vspace{2mm}
 \\ 
+p_\nu \Psi ^{[\mu \nu ]}-(p^\mu -\gamma _\lambda p^\lambda \gamma ^\mu
)\gamma _\lambda \gamma _\sigma \Psi ^{[\lambda \sigma ]}=0,
\end{array}
\label{4.9}
\end{equation}
in which
 $f$ denotes the expression $p_\lambda \Psi ^\lambda -\gamma _\lambda p^\lambda \gamma
_\nu \Psi ^\nu .$

Reducing (\ref{4.8}) with $\gamma _\mu \gamma _\nu $ we obtain condition (%
\ref{4.2}). Thus equations (\ref{4.8}) and  (\ref{4.9}) can be simplified to 
\begin{equation}
\begin{array}{l}
(\gamma _\lambda p^\lambda -m)\Psi ^{[\mu \nu ]}-p^\mu \Psi ^\nu +p^\nu \Psi
^\mu \\ 
+\frac 1{12}(\gamma ^\mu \gamma ^\nu -\gamma ^\nu \gamma ^\mu )(f-2p_\lambda
\gamma _\sigma \Psi ^{[\lambda \sigma ]})=0,
\end{array}
\label{4.10}
\end{equation}
and
\begin{equation}
\begin{array}{l}
(\gamma _\lambda p^\lambda +m)\Psi ^\mu -\gamma ^\mu (f+m\gamma _\lambda
\Psi ^\lambda )
-p^\mu \gamma _\lambda \Psi ^\lambda +p_\nu \Psi ^{[\mu \nu
]}=0.  \label{4.11}
\end{array}
\end{equation}
respectively.

Reducing equations (\ref{4.10}) successively with $\gamma _\mu $, $p_\mu $ and  $p_\mu
\gamma _\nu $, and (\ref{4.11}) with $\gamma _\mu $ and $p_\mu $, we obtain
equations (\ref{4.1})-(\ref{4.3}) for $\Psi ^{]\mu \nu] }$ and the condition $%
\Psi ^\mu =0.$ In other words, the equations of motion annul the auxiliary
field $\Psi ^\mu$ and are equivalent to the system (\ref{4.1})-(\ref{4.3}) describing a
particle of spin $\frac 32$ and mass $m$.

Taking into account relation (\ref{4.2}) it is convenient to represent $\Psi
^{[\mu \nu ]}$ in the form 
\begin{equation}
\Psi ^{[\mu \nu ]}=\chi ^{\mu \nu }+\frac 12(\gamma ^\mu A^\nu -\gamma ^\nu
A^\mu )  \label{9.3},
\end{equation}
where $\chi ^{\mu \nu }$and $A^\mu $ is a $\gamma $-irreducible tensor and
vector, respectively. They satisfy the conditions: $\chi ^{\mu \nu
}=-\chi ^{\nu \mu },\,\gamma _\nu \chi ^{\mu \nu }=0$ and $\gamma _\nu A^\nu =0.$
In view of (\ref{4.2}) we easily find that 
$
\chi ^{\mu \nu }=\frac 12\Psi _{+}^{[\mu \nu ]}$ and $\quad A^\mu =-\gamma _\nu
\Psi ^{[\mu \nu ]}$.

Using variables (\ref{9.3}) and introducing a minimal interaction via
replacement $p_\mu$ by $ \pi _\mu $ we can write the related
equations (\ref{4.2}), (\ref{4.10}) and  (\ref{4.11}) in the following
equivalent form 
\begin{equation}
\begin{array}{l}
\pi ^\mu (\Psi ^\nu -A^\nu )-\pi ^\nu (\Psi ^\mu -A^\mu )+\frac 1{12}(\gamma
^\mu \gamma ^\nu -\gamma ^\nu \gamma ^\mu )\widehat{f} \\ 
+m\left( \chi ^{\mu \nu }-\frac 12(\gamma ^\mu A^\nu -\gamma ^\nu A^\mu
)+\gamma ^\mu \Psi ^\nu -\gamma ^\nu \Psi ^\mu \right) =0,
\end{array}
\label{9.4}
\end{equation}
\begin{equation}
2\gamma _\lambda \pi ^\lambda \chi ^{\mu \nu }-m(\gamma ^\mu A^\nu -\gamma
^\nu A^\mu )-(\pi \cdot \Psi )_{+}^{\mu \nu }=0,  \label{9.5}
\end{equation}
where the following  notations have been used: \[\ba{l}(\pi \cdot \Psi )_{\pm }^{\mu \nu }
=\pi ^\mu \Psi ^\nu -\pi
^\nu \Psi ^\mu \pm \frac 12\gamma _5\varepsilon^{\mu \nu \rho \sigma }\pi
_\rho \Psi _\sigma ,
\vspace{2mm}
 \\  \widehat{f}=\pi _\lambda \Psi ^\lambda -\gamma
_\lambda \pi ^\lambda \gamma _\nu \Psi ^\nu .\ea\]

We show in Appendix B  that for a rather extended class of external fields
equations (\ref{9.4}) and (\ref{9.5}) remain causal.

\section{Discussion}

In the paper  RWE for a massive interacting particle  with arbitrary half integer
spin $s$ has been proposed and especially for  $s=3/2$ discussed in detail.

RWE considered in Section 3 are causal and free of most inconsistences which are
typical for equations for particles of spin greater then $1$. Moreover,
these equations have a physically suitable  form in  quasirelativistic approximation and
are able to describe mostly used interactions such as Pauli, spin-orbit, quadrupole and Darwin
couplings. We remind that even such popular equation as the
Kemmer-Duffin-Petiau \cite{40} one does not describe the spin-orbit coupling
in the framewoork  of the minimal interaction principle \cite{23A}.

The other attractive feature of the tensor-spinorial wave equations consists
in their hidden simplicity which can be recognized considering the
second-order equation (\ref{3.19}) for the physical components. This
equation can be easily solved for many particular cases of the external
fields like it was done in \cite{40}, \cite{41} for the special case of $g={%
\frac 1s}$. We plan to present these exact solutions elsewhere.

The considered equations have a reasonable zero mass limit for a free particle case and so can serve
as a basis to formulate consistent equations for massless fields with
arbitrary spin. Such equations were discussed briefly in Section VI.

Finally, introduction of anomalous interaction into the tensor-spinorial
wave equations generates a surprisingly small complexity of the theory
in comparison with the case of the minimal interaction. In this aspect the
proposed equations are quite unique and are  more convenient than
even the Dirac equation!

We do not discuss specific kind of difficulties connected with the complex
energy eigenvalues for the case of interaction with the constant magnetic
field provided the gyromagnetic ratio $g$ of the particle is equal to 2 \cite
{20}. This problem arises also for the tensor-spinorial wave
equation, but it can be overcome using the approach proposed in \cite{42}.

For completeness notice that single particle equations for spin ${\frac 32}$
considered in Section VII correspond to the Harish-Chandra index 4 and thus
belong to the class described by Labont\'e \cite{43}. We believe that our tensor-spinorial
formulation (\ref
{4.5})-(\ref{4.11}) and (\ref{9.4}), (\ref{9.5})  forms an appropriate basis for the theory
of interacting particles of arbitrary
half-integer spin and its various applications. 
\renewcommand{\theequation}{A\arabic{equation}} %
\setcounter{equation}{0} \appendix

\section{Inconsistency of Singh-Hagen equations}

A specific formulation of the RS equations was used by Singh and Hagen \cite
{8} who introduced an additional scalar-bispinor field $\psi $ such that $%
\psi ^\mu $ and $\psi $ satisfy the following system 
\begin{equation}
\begin{array}{l}
\widetilde{{\cal F}}^\mu =\left( \gamma ^\nu p_\nu +m\right) \widetilde{\psi 
}^\mu -{\frac 12}\gamma ^\mu p_\lambda \widetilde{\psi }^\lambda 
-{\frac 23}\left( p^\mu -{\frac 14}\gamma ^\mu \gamma ^\nu p_\nu \right) 
\widetilde{\psi }=0,
\vspace{2mm}
 \\ 
\widetilde{{\cal F}}=p_\nu \widetilde{\psi }^\nu -\left( \gamma ^\nu p_\nu
-2m\right) \widetilde{\psi }=0,
\vspace{2mm}
 \\  \gamma _\mu \widetilde{\psi }^\mu =0.
\end{array}
\label{sing}
\end{equation}

Equations (A1) are equivalent to the RS equations. Indeed,
denoting in (2.1) $ \widetilde{\psi }^\mu +\frac 13\gamma ^\mu 
\widetilde{\psi }$  by $\psi ^\mu$ we easily find that (A1) is an algebraic consequence of
(2.1) and vice versa, because 
\begin{equation}
\begin{array}{l}
\widetilde{{\cal F}}=\frac 12\gamma _\mu {\cal F}^\mu ,
\widetilde{%
{\cal F}}^\mu ={\cal F}^\mu -\frac 14\gamma ^\mu \gamma _\lambda {\cal F}%
^\lambda , 
\vspace{2mm}
 \\  
{\cal F}^\mu =\widetilde{{\cal F}}^\mu +\frac 12\gamma ^\mu \widetilde{{\cal %
F}}.
\end{array}
\label{A.2}
\end{equation}

In contadistinction to the RS equation, it was stated in \cite{27} that the Singh-Hagen
formulation (A1)  is causal provided a nontrivial
anomalous interaction is introduced. We think that this statement has no meaning
since in the case of anomalous interaction proposed in \cite{27} 
the Singh-Hagen equations became inconsistent. This can be easily seen in the following way.
The equation proposed in \cite{27} has
the form 
\begin{equation}
\begin{array}{l}
\widetilde{{\cal F}}^\mu =\left( \gamma ^\nu p_\nu +m\right) \widetilde{\psi 
}^\mu -{\frac 12}\gamma ^\mu p_\lambda \widetilde{\psi }^\lambda  
-{\frac 23}\left( p^\mu -{\frac 14}\gamma ^\mu \gamma ^\nu p_\nu \right) 
\widetilde{\psi }+\alpha F_{+}^{\mu \nu }\widetilde{\psi }_\nu =0,
\vspace{2mm}
 \\ 
\widetilde{{\cal F}}=p_\nu \widetilde{\psi }^\nu -\left( \gamma ^\nu p_\nu
-2m\right) \widetilde{\psi }=0,
\vspace{2mm}
 \\ \gamma _\mu \widetilde{\psi }^\mu =0
\end{array}
\label{AA3}
\end{equation}
where $\alpha $ is a coupling constant.

Using relations (\ref{A.2}) we reduce (\ref{AA3}) to the RS form: 
\begin{equation}
\begin{array}{l}
\left( \gamma ^\nu \pi _\nu +m\right) \psi ^\mu -\gamma ^\mu \pi _\alpha
\psi ^\alpha -\pi ^\mu \gamma _\alpha \psi ^\alpha  
+\gamma ^\mu \left( \gamma ^\nu \pi _\nu -m\right) \gamma _\lambda \psi
^\lambda +T^{\mu \nu }\psi _\nu =0
\end{array}
\label{RA}
\end{equation}
where 
\begin{equation}
T^{\mu \nu }=\alpha \left( F_{+}^{\mu \nu }-\frac 14F_{+}^{\mu \lambda
}\gamma _\lambda \gamma ^\nu \right) .  \label{tensor}
\end{equation}

It is easy to show that in contrast with (\ref{2}), equation (\ref{RA}) does
not include required eight constraints but only four of them. Indeed, reducing (%
\ref{RA}) with $\gamma _\mu $ and $\pi _\mu $we obtain the correct number of
constraints only for the case $T^{00}=0$ \cite{38}, which is compatible with
(\ref{tensor}) only for the trivial anomalous interaction $\alpha F_{+}^{\mu
\nu }=0.$


\section{Consistency of equations for singlets}

\renewcommand{\theequation}{B\arabic{equation}} \setcounter{equation}{0} Let
us show that for some class of external fields equations (\ref{9.4}), (\ref
{9.5}) are consistent, i.e., include the correct number of constraints and
are hyperbolic. To do this we will use also differential and algebraic
consequences of these equations.

Contracting (\ref{9.4}), (\ref{9.5}) with $\gamma _\nu $ and $\gamma _\mu
\gamma _\nu $  and using (\ref
{9.4}), (\ref{9.5}) we come to the equivalent $\gamma $-irreducible set of equations: 
\begin{equation}
\begin{array}{l}
(\pi \cdot \Psi )_{-}^{\mu \nu }-(\pi \cdot A)_{-}^{\mu \nu }+2m\chi ^{\mu
\nu }  
+\frac 16(\gamma ^\mu \gamma ^\nu -\gamma ^\nu \gamma ^\mu )(\widehat{f}%
+3m\gamma _\lambda \Psi ^\lambda )=0,
\end{array}
\label{9.6}
\end{equation}
\begin{equation}
\begin{array}{l}
\gamma _\lambda \pi ^\lambda (\Psi ^\mu -A^\mu )-\pi ^\mu \gamma _\lambda
\Psi ^\lambda 
+m(2\Psi ^\mu -A^\mu )+\gamma ^\mu (\frac 12\widehat{f}+m\gamma _\lambda
\Psi ^\lambda )=0,
\end{array}
\label{9.7}
\end{equation}
\begin{equation}
\begin{array}{l}
2\pi _\nu \chi ^{\mu \nu }+\gamma _\lambda \pi ^\lambda A^\mu -2m(\Psi ^\mu
-A^\mu )-  
\gamma ^\mu (\widehat{f}+m\gamma _\lambda \Psi ^\lambda )=0,
\end{array}
\label{9.8}
\end{equation}
\begin{equation}
\pi _\nu A^\nu -2\widehat{f}-3m\gamma _\lambda \Psi ^\lambda =0.  \label{9.9}
\end{equation}

The other (differential) consequences can be found by reducing (\ref{9.6})-(%
\ref{9.8}) with $p_{\mu }.$ In this way we obtain from (\ref{9.7}), (\ref
{9.8}) the following two relations: 
\begin{equation}
\gamma _{\lambda }\pi ^{\lambda }\widehat{f}+6m^{2}\gamma _{\lambda }\Psi
^{\lambda }-2i\gamma _{\lambda }\widetilde{F}^{\lambda \sigma }(\Psi
_{\sigma }-A_{\sigma })=0  \label{9.10}
\end{equation}
and 
\begin{equation}
m\widehat{f}=i\left( \gamma _{\lambda }\widetilde{F}^{\lambda \sigma }(\frac{%
1}{2}A_{\sigma }-\Psi _{\sigma })+\frac{1}{2}\widetilde{F}_{\lambda \sigma
}\chi ^{\lambda \sigma }\right) .  \label{9.11}
\end{equation}

One more consequence can be obtained reducing (\ref{9.6}) with $\pi _\nu ,$
acting on (\ref{9.7}) by $\gamma _\lambda \pi ^\lambda $ and adding the
resultant expressions together. We get 
\begin{equation}
m^2\Psi ^\mu +\frac 16\pi ^\mu \widehat{f}+i\left( \widetilde{F}^{\mu \sigma
}-\frac 13\gamma ^\mu \gamma _\lambda \widetilde{F}^{\lambda \sigma }\right)
(\Psi _\sigma -A_\sigma )=0.  \label{9.12}
\end{equation}

Reducing (\ref{9.12}) once more with $\pi _\mu $ we obtain a scalar
consequence 
\begin{equation}
\left( m^2-\frac i{12}\gamma _\mu \gamma _\nu F^{\mu \nu }\right) \widehat{f}%
=-i\pi _\lambda \tilde{F}^{\lambda \sigma }(\Psi _\sigma -A_\sigma ).
\label{9.121}
\end{equation}

Applying operator $\gamma _\lambda \pi ^\lambda $ to (\ref{9.12}) and using (%
\ref{9.7}), (\ref{9.10}) we come to one more consequence 
\begin{equation}
\begin{array}{l}
m^2(\gamma _\lambda \pi ^\lambda \Psi ^\mu -\frac i3\pi ^\mu \gamma _\lambda
\Psi ^\lambda )
+\frac i3(\gamma ^\mu \gamma _\lambda \pi ^\lambda -\pi ^\mu )\gamma
_\lambda \widetilde{F}^{\lambda \sigma }(\Psi _\sigma -A_\sigma ) \\ 
-i\gamma _\lambda \pi ^\lambda \widetilde{F}^{\mu \sigma }(\Psi _\sigma
-A_\sigma )+\frac i6F^{\mu \lambda }\gamma _\lambda \widehat{f}=0,
\end{array}
\label{9.13}
\end{equation}

Finally, contracting (\ref{9.12}) with $F^{\mu \lambda }\pi _\lambda $ and
using (\ref{9.5}) we get the following important condition  
\begin{equation}
\begin{array}{l}
m^2F^{\mu \lambda }\pi _\lambda \Psi ^\mu -\frac i3F^{\mu \lambda }\pi
_\lambda \gamma ^\mu \gamma _\lambda \widetilde{F}^{\lambda \sigma }(\Psi
_\sigma -A_\sigma )  
-i\gamma _5C_2(\pi _\sigma \Psi ^\sigma -2\widehat{f}\\-3m\gamma _\lambda \Psi
^\lambda )  
-F^{\mu \lambda }\left( \frac \partial {\partial x^\lambda }\widetilde{F}%
^{\mu \sigma }\right) (\Psi _\sigma -A_\sigma )=0,
\end{array}
\label{9.14}
\end{equation}
where $C_2=F^{\mu \lambda }\widetilde{F}_{\mu \lambda }$ is an invariant of
the electromagnetic field.

Now we are ready to analyze the constraint context of equations (\ref{9.4}) and 
(\ref{9.5}). First we note that the considered system includes nine dependent
variables (each being a four-component spinor). To describe a particle of
spin $\frac 32$ it is sufficient to have eight degrees of freedom and so we
need seven constraints which do exist. Six of them are presented by equations (%
\ref{9.4}) for $\mu ,\nu =1,2,3$ and (\ref{9.12}) for $\mu =1,2,3.$ The
seventh constraint is easily obtained from (\ref{9.8}) for $\mu =0$ and (\ref
{9.9}): 
\[
\begin{array}{l}
2\pi _a\chi ^{0a}-\gamma _a\pi _aA^0+\gamma ^0(\pi _aA_a+2m\gamma _\lambda
\Psi ^\lambda +\widehat{f})  
+2m(\Psi ^0-A^0)=0.
\end{array}
\]

The next task is to find the true motion equations. They are given by equation (\ref
{9.4}) for $\mu =1,2,3,\nu =0,$ by equations (\ref{9.8}), (\ref{9.13}) for $\mu
=1,2,3$ and by equation (\ref{9.14}). The related matrix with time derivatives
is non-singular provided 
\begin{equation}
C_2\neq 0\text{ or (and) }\widetilde{F}^{0a}\neq 0.  \label{9.15}
\end{equation}

On the other hand, if $C_2=0$ and $\widetilde{F}^{0a}=0$ then relation (\ref
{9.12}) for $\mu =0$ reduces to the constraint expressing $\psi ^0$ via
other variables and so that in this case we do not need a motion equation for $%
\psi ^0.$

To investigate causality we consider the true motion equations in the eikonal
approximation. Substituting the characteristic four-vector $n_\mu $ to the
covariant derivatives and keep only leading terms in $n_\mu $ we come to the
following system 
\begin{equation}\begin{array}{l}
n^\mu (\Psi ^\nu -A^\nu )-n^\nu (\Psi ^\mu -A^\mu )=0, 

\vspace{2mm}

\\2n_\nu \chi ^{\mu \nu }+\gamma _\lambda n^\lambda A^\mu =0,  

\vspace{2mm}

\\m^2(\gamma _\lambda n^\lambda \Psi ^\mu -n^\mu \gamma _\lambda \Psi ^\lambda
) 
+\frac i3(\gamma ^\mu \gamma _\lambda n^\lambda -n^\mu )\gamma _\lambda 
\widetilde{F}^{\lambda \sigma }(\Psi _\sigma -A_\sigma )=0,

\vspace{2mm}

\\m^2F^{\mu \lambda }n_\lambda \Psi _\mu -\frac i3F^{\mu \lambda }n_\lambda
\gamma _\mu \gamma _\alpha \widetilde{F}^{\alpha \sigma }(\Psi _\sigma
-A_\sigma ) 
-i\gamma _5C_2n_\sigma \Psi ^\sigma =0.
\end{array}
\label{9.19}
\end{equation}
Setting $n_\mu =(n,0,0,0)$ in (\ref{9.19}) we easily find that $%
\chi ^{\mu \nu }=A^\mu =\Psi ^\mu =0$ provided $n_0$ and $C_2$ are not equal to zero.
Thus equations (\ref{9.4}), (\ref{9.5}) are causal provided the external
electromagnetic field satisfies the covariant relation $C_2\neq 0.$  

We remind that acausality of the RS equation is caused by non-covariance of its 
hyperbolicity condition  \cite{44}.%

\end{document}